\documentclass[preprint2,10.5pt]{aastex}
\begin{document}

\title{\large{\rm{DISCOVERY OF THE HOST CLUSTER FOR THE FUNDAMENTAL \\ CEPHEID CALIBRATOR $\zeta$ GEM}}}
\author{D. Majaess$^{1,4}$, D. Turner$^{1,4,5,6}$, W. Gieren$^2$, D. Balam$^3$, D. Lane$^{1,4}$}
\affil{$^1$ Department of Astronomy \& Physics, Saint Mary's University, Halifax, NS, Canada}
\affil{$^2$ Departamento de Astronom\'{i}a, Universidad de Concepci\'on, Concepci\'on, Chile.}
\affil{$^3$ Dominion Astrophysical Observatory, Victoria, BC, Canada.}
\affil{$^4$ Abbey Ridge Observatory, Halifax, Nova Scotia, Canada.}
\affil{$^5$ Visiting Astronomer, David Dunlap Observatory.}
\affil{$^6$ Visiting Astronomer, Kitt Peak National Observatory.}
\email{dmajaess@cygnus.smu.ca}
\begin{abstract}
New and existing CORAVEL, $UBVJHK_s$, HST, HIP/Tycho, ARO, KPNO, and DAO observations imply that the fundamental Cepheid calibrator $\zeta$ Gem is a cluster member.  The following parameters were inferred for $\zeta$ Gem from cluster membership and are tied to new spectral classifications (DAO) established for 26 nearby stars (e.g.,~HD53588/B7.5IV, HD54692/B9.5IV): $E_{B-V}=0.02\pm 0.02$, $\log{\tau}=7.85\pm0.15$, and $d=355\pm15$ pc.   The mean distance to $\zeta$ Gem from cluster membership and six recent estimates (e.g., IRSB) is $d=363\pm 9(\sigma_{\bar{x}}) \pm 26 (\sigma )$ pc.   The results presented here support the color-excess and HST parallax derived for the Cepheid by \citet{be07}.  Forthcoming precise proper motions (DASCH) and Chandra/XMM-Newton observations of the broader field may be employed to identify cluster members, bolster the cluster's existence, and provide stronger constraints on the Cepheid's fundamental parameters. 
\end{abstract}
\keywords{Hertzsprung-Russell and C-M diagrams, open clusters and associations, stars: distances, stars: variables: Cepheids}

\section{{\rm \footnotesize INTRODUCTION}}
An independent distance determination for $\zeta$ Gem is desirable since HST and HIP parallaxes for the classical Cepheid exhibit an unsatisfactory spread: $d=358-422$ pc \citep{pe97,vl07,be07}.  Establishing reliable parameters for $\zeta$ Gem is particularly important given the Carnegie Hubble and S$H_0$ES projects \citep{mr09,fm10} are relying on HST calibrators \citep{be07} to break degeneracies hindering the selection of a cosmological model \citep{ri11}. The Carnegie Hubble project shall likewise employ Galactic calibrators tied to open clusters \citep{tu10} featuring IRSB corroborated distances \citep{gi05,st11}. The classical Cepheid $\zeta$ Gem is the second longest period calibrator possessing a precise HST parallax \citep{be02,be07}, and a solid calibration for such stars is needed given longer-period Cepheids may be less affected by (insidious) photometric contamination \citep[][their Fig.~17]{mac06}.   Moreover, longer-period Cepheids are detectable in distant galaxies owing to their increased luminosity relative to shorter-period Cepheids \citep[e.g.,][their Fig.~1]{gi09}.  Sampling remote galaxies in the Hubble flow mitigates uncertainties tied to peculiar velocity corrections and hence $H_0$ \citep[][their Fig.~1]{fr01}.  

In this study, CORAVEL, $UBVJHK_s$, HST, HIP/Tycho, KPNO and DAO spectroscopic observations are employed to identify stars potentially associated with $\zeta$ Gem, thereby permitting its fundamental properties to be deduced from cluster membership ($\log{\tau}$, $E_{B-V}$, $W_{VI_c,0}$).  %

\section{{\rm \footnotesize ANALYSIS}}
HIP/Tycho data were examined for stars surrounding $\zeta$ Gem which exhibit similar proper motions: $-10<\mu_{\alpha}<-2$ and $-7<\mu_{\delta}<2$ \citep{pe97,vl07}.  Stars which display proper motions that deviate significantly from $\zeta$ Gem were eliminated via the aforementioned limits.  However, countless field stars likewise exhibit proper motions marginally offset from zero.  The available proper motions could not be employed to substantiate the cluster's existence given their uncertainties \citep[][see their Fig.~3 for $\delta$ Cep]{ma12}.  Stars redder than $B-V\sim0.14$ were culled to further mitigate field contamination (e.g., red clump giants).  $UBVJHK_s$ photometry was tabulated for the remaining sample using the compilations of \citet{me91}, \citet{pe97}, and \citet{cu03}.

\begin{figure}[!t]
\epsscale{0.8}
\plotone{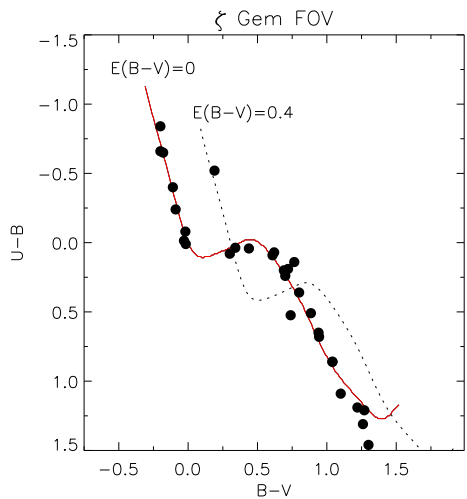}
\caption{\small{Color-color diagram for all stars $r\le4 \degr$ from $\zeta$ Gem which possess $UBV$ photometry \citep{me91}.  The field is comparatively unreddened.  The intrinsic $UBV$ relation and reddening law for the region were adopted from \citet{tu76,tu89}.  Most objects featured in the diagram are unassociated with $\zeta$ Gem.}}
\label{fig-ubv}
\end{figure}

\subsection{{\rm \footnotesize REDDENING AND AGE}}
\label{s-age}
\begin{figure*}[!t]
\epsscale{1.3}
\plotone{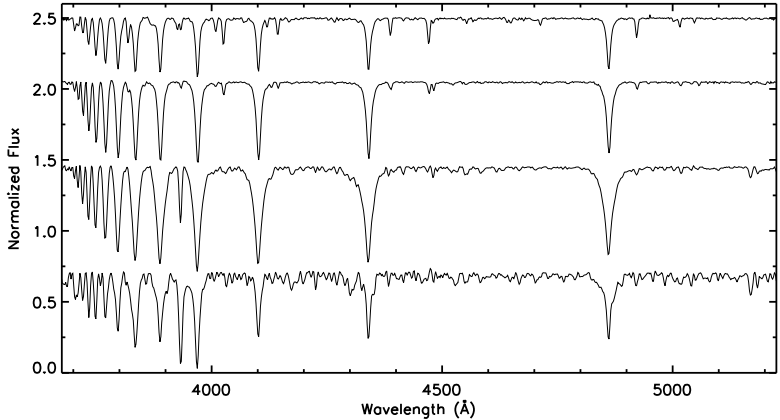}
\caption{\small{A subset of the new DAO spectra obtained for objects in the field of the classical Cepheid $\zeta$ Gem.  From top to bottom the spectra correspond to HD50767, HD53588, HD52474, and $\zeta$ Gem B.}}
\label{fig-sp}
\end{figure*}

An analysis of \textit{all} stars near $\zeta$ Gem featuring $UBV$ photometry \citep{me91} confirms that the field is comparatively unreddened (Fig.~\ref{fig-ubv}).   The mean color-excess inferred from new DAO spectra (Fig.~\ref{fig-sp}) for probable cluster members in Table~\ref{table1} is $E(B-V)=0.019\pm 0.017$.   The findings support the reddening determined for $\zeta$ Gem by \citet{be07}. 

The cluster age can be constrained by examining the spectral type of members near the turnoff.  A distinct sequence of B-stars is visible in the $BVJH$ color-magnitude diagrams (Fig.~\ref{fig-cmd}, $r\le 2.5\degr$ from $\zeta$ Gem). The stars HD51102, HD51353, HD53588, and HD55919 aggregate near the turnoff and exhibit B6-B7 spectral classes (Table~\ref{table1}).  For example, HD53588 displays $UBV$ colors and a DAO spectrum conducive to a B7.5 IV (Table~\ref{table1}).  The star's radial velocity ($RV=10\pm3$ km/s) is consistent with that established for $\zeta$ Gem ($RV\sim7$ km/s).  However, cluster membership cannot be established solely on the basis of consistent radial velocities since the predicted $RV$-distance gradient along $\ell\sim197 \degr$ is shallow.  Membership for the two earliest type stars examined is less certain (HD50767, HD51354, Table~\ref{table1}).  A spectroscopic parallax for HD50767 implies that the star is well behind the cluster.  The spectroscopic and HIP parallaxes for HD50767 disagree.  The intrinsic colors for HD51354 ($(B-V)_0:(U-B)_0=-0.18:-0.65$) and the DAO spectrum indicate that the star is a B3 Vnne (H$\beta$ emission).  A mean of the HIP parallaxes established for HD51354 (Table~\ref{table1}) is consistent with that established for $\zeta$ Gem \citep[$\pi=2.78\pm0.18$ mas,][]{be07}.  The \citet{pe97} parallax for HD51354 is $\pi=2.57\pm0.81$ mas, whereas the revised HIP parallax is $\pi=3.75\pm0.47$ mas \citep{vl07}.  Spectroscopic parallaxes for emission stars (i.e., HD51354) are unreliable.

A $\log{\tau}=7.85\pm0.15$ Padova isochrone\footnote{http://pleiadi.pd.astro.it/} provides the best match to the evolutionary track of the bluer members, $\zeta$ Gem, and is consistent with the age inferred from members aggregating near the turnoff (Fig.~\ref{fig-cmd}). The age likewise agrees with that predicted for $\zeta$ Gem \citep{tu96,bo05}.

\subsection{{\rm \footnotesize LOWER-MASS CLUSTER MEMBERS}}
\label{s-lm}
\begin{figure*}[!t]
\epsscale{1.2}
\plotone{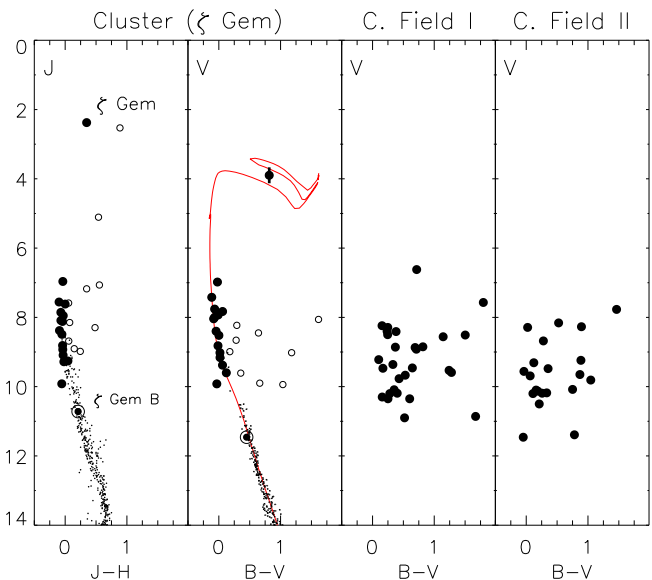}
\caption{\small{Left to right, $JHBV$ color-magnitude diagrams for HIP/Tycho stars within $r \le 2.5 \degr$ of $\zeta$ Gem featuring $-10<\mu_{\alpha}<-2$ and $-7<\mu_{\delta}<2$.   An early type cluster sequence is absent from the comparison fields (panels 3, 4, HIP/Tycho), which encompass $r\le4 \degr$. \emph{Small dots} denote calibration stars from \citet{ma11}, which were employed to tie the cluster distance to a geometrically anchored scale \citep{vl09,ma11}. Open circles are likely field stars.  Panel 2, a Padova $\log{\tau}=7.85$ isochrone was applied.  $\zeta$ Gem and $\zeta$ Gem B are the brightest cluster member (amplitude variation indicated) and circled dot, accordingly.}}
\label{fig-cmd}
\end{figure*}

\citet{tu78} noted that $\zeta$ Gem B (J2000 07:04:12.73 +20:34:21.3) may be associated with $\zeta$ Gem. The stars are separated by $r \sim1.4\arcmin$.   A mean radial velocity was determined for $\zeta$ Gem B from six CORAVEL measurements obtained at l'Observatoire de Haut-Provence (OHP).\footnote{M. Mayor kindly obtained CORAVEL observations of $\zeta$ Gem B for D. Turner.}  The radial velocities acquired from the OHP span -16.2 to 36.9 km/s, yielding a mean of 9.9 km/s.  $\zeta$ Gem B is thus a spectroscopic binary since the uncertainty tied to an individual measurement is 1.4 km/s.  The mean radial velocity matches that established for the Cepheid, to within the uncertainties.  However, as noted in \S \ref{s-age}, the shallow radial velocity-distance gradient along the line of sight requires that cluster membership be secured by independent means.  $UBV$ photometry by \citet{fe69} indicates that $\zeta$ Gem B is an F-type star \citep{tu78}.  That is corroborated by the 2MASS colors for the object ($(J-H):(H-K)=0.206:0.056$), which are indicative of an unreddened F5-F8V \citep{sl09}.   A spectrogram of the star was acquired on HJD=2444122.995 from the 2.1-m telescope at Kitt Peak.  That spectrogram and a DAO spectrum confirm that $\zeta$ Gem B is an F4V (Table~\ref{table1}, Fig.~\ref{fig-sp}).  The result is supported by \citet{be07}, who classified the star as an F3.5V.  The spectral and luminosity class for $\zeta$ Gem B are consistent with that expected for a cluster member at the star's location in the $BVJH$ color-magnitude diagrams (Fig.~\ref{fig-cmd}).

DAO spectra were obtained for two additional stars in close proximity to $\zeta$ Gem ($r<6 \arcmin$).  The spectral and luminosity class for 2MASS 07041267+2030196 are consistent with that expected for a cluster member at the star's location in the $BVJH$ color-magnitude diagrams (Fig.~\ref{fig-cmd}, Table~\ref{table1}).  The same is true for 2MASS 07035262+2035162. \citet{be07} inferred an analogous classification for 2MASS 07035262+2035162 (F6V).  Four additional stars in close proximity to $\zeta$ Gem exhibit multiband photometry conducive to late-type (potential) cluster members.  New $BV$ observations for those stars were acquired from the Abbey Ridge Observatory \citep[ARO,][]{la08,ma08}. The data were processed via ARAP \citep{la08} and DAOPHOT \citep{st87}, and subsequently standardized to photometry obtained from the New Mexico State University 1-m telescope (T. Harrison priv.~communication, see also \citealt{be07}).  The following equations were derived to place the instrumental ARO photometry onto the Johnson system: 
\begin{eqnarray}
\nonumber
B-V=(0.99\pm0.06)\times(b-v)-(0.57\pm0.08) \\
\nonumber
V-v=-0.86\pm0.02 
\end{eqnarray}
Observations from the AAVSO's Bright Star Monitor (BSM) provided additional data for $\zeta$ Gem B and 2MASS 07041267+2030196.   The BSM is located at the Astrokolkhoz telescope facility near Cloudcroft, New Mexico.

HST WFC3 images \citep[HST Proposal 12215,][]{ev09} reveal a star $15 \arcsec$ west of the Cepheid at J2000 07:04:05.5 +20:34:12.0.   The object was identified after subtracting (normalized) the image featuring $\zeta$ Gem from a master, which was constructed (median combine) using Cepheids observed for proposal 12215. The object displays a signal-noise-ratio greater than 20 in both WFC3 images.  However, N. Evans (priv.~communication) noted that the star's photometry is not conducive to cluster membership.  

Later-type stars associated with $\zeta$ Gem should presumably exhibit x-ray emission \citep{ev11}.  Additional observations are required to assess the candidates (e.g., Chandra/XMM-Newton).   The objects are not featured in the ROSAT catalogs.

\subsection{{\rm \footnotesize CLUSTER DISTANCE}}
\label{s-distance}
A precise cluster distance may be determined since the reddening, age, and chemical composition of the sample are constrained \citep[${\rm [ Fe/H]}_{\rm \zeta Gem }\sim0$,][]{lu11}.  The magnitude shift required to overlay the intrinsic relation upon the data yields $d=355\pm15$ pc (Fig.~\ref{fig-cmd}).   The zero-point is tied to seven benchmark open clusters ($d<250$ pc) which exhibit matching $JHK_s$ and revised HIP distances \citep[e.g., the Hyades, $\alpha$ Per, Coma Ber,][]{vl09,ma11}.  A redetermination of the HST parallax for the Hyades supports that scale \citep{mc11}.  The scale employed here is anchored to clusters where consensus exists, rather than the discrepant case (i.e., the Pleiades).  A ratio of total to selective extinction $R_J$ was adopted from \citet{ma11b} \citep[see also][]{bo04}, whereas a value for $R_V$ was adopted from \citet{tu76}.  Deviations from the canonical reddening law are mitigated since the field is comparatively unreddened (Fig.~\ref{fig-ubv}, note that $V_0=V-E_{B-V} \times R_V$).

A mean distance inferred from potential members possessing revised HIP parallaxes is $d=366\pm 57(\sigma_{\bar{x}}) \pm 196 (\sigma )$ pc.  Certain stars were excluded from the derivation since they exhibit uncertainties greater than the parallax cited, deviate significantly from the mean, or feature negative parallaxes (e.g., HD49824, $\pi=-0.74\pm0.89$ mas).  Stars with spectral types inconsistent with cluster membership were likewise removed (e.g., HD50767, \S \ref{s-age}).  

Three HIP parallaxes exist for $\zeta$ Gem: $\pi=2.79\pm0.81:2.37\pm0.30:2.71\pm0.17$ mas \citep{pe97,vle07,vl07}. The results inferred from cluster membership and HST observations for $\zeta$ Gem \citep[$\pi=2.78\pm0.18$ mas,][]{be07} support the \citet{vl07} HIP determination.  Distances to $\zeta$ Gem are also available from the infrared surface brightness technique \citep[$d=386\pm9$ pc,][]{st11} and the PTI \citep[$d=336\pm44$ pc,][and references therein]{la00}.  A straight mean of the 6 (+2 cluster based) estimates to $\zeta$ Gem yields $d=363\pm 9(\sigma_{\bar{x}}) \pm 26 (\sigma )$ pc.  A weighted mean of $d=368$ pc was obtained by assigning w=1 for the HIP and PTI estimates, and w=2 for the rest.   

\section{{\rm \footnotesize CONCLUSION \& FUTURE RESEARCH}}
The evidence indicates that the 10$^{\rm d}$ classical Cepheid calibrator $\zeta$ Gem belongs to a newly identified cluster (Fig.~\ref{fig-cmd}).  Potential members were identified on the basis of new and existing CORAVEL, $UBVJHK_s$, HST, HIP/Tycho, KPNO and DAO ($n=26$) spectroscopic observations.  The analysis indicates that the field is comparatively unreddened (Fig.~\ref{fig-ubv}), and that $\zeta$ Gem exhibits $E(B-V)=0.019\pm0.017$ (\S \ref{s-age}).  The age and distance were inferred from spectroscopic and multiband $UBVJHK_s$ analyses of the cluster stars, yielding $\log{\tau}=7.85\pm0.15$ and $d=355\pm15$ pc (Fig.~\ref{fig-cmd}).  The results  support the parameters derived for $\zeta$ Gem by \citet{be07}.  However, the breadth of the standard deviation tied to several recent estimates for $\zeta$ Gem is unsatisfactory (\S \ref{s-distance}, $d=363\pm 9(\sigma_{\bar{x}}) \pm 26 (\sigma )$), and continued research is required.  CORAVEL, KPNO, and DAO observations indicate that $\zeta$ Gem B is an F-type spectroscopic binary.  DAO spectra were likewise obtained for two other low-mass members in close proximity to $\zeta$ Gem.  Additional observations are required to highlight \textit{bona fide} members on the candidate list (Table~\ref{table1}).  Those stars are potential members pending further evidence.  Membership identification may be facilitated by obtaining precise proper motions for fainter stars near $\zeta$ Gem from photographic plates stored at the CfA \citep[][DASCH]{gr07}.\footnote{Digital Access to a Sky Century @ Harvard (DASCH), \url{http://hea-www.harvard.edu/DASCH/}}  The plates offer multi-epoch coverage spanning a $\sim100$ year baseline, and uncertainties are further mitigated owing to sizable statistics ($\sim (5-10) \times 10^2$ plates per object).  Searching for x-ray emission from lower-mass stars near $\zeta$ Gem is likewise a viable pursuit for corroborating membership \citep{ev11,ev11b}.   However, the star Polaris B is of a similar spectral type as $\zeta$ Gem B \citep{tu77}, and does not exhibit x-ray emission \citep{ev10}.

\begin{deluxetable}{crrrrrrcrr}
\tablewidth{0pt}
\tabletypesize{\scriptsize}
\tablecaption{Stars Near $\zeta$ Gem}
\tablehead{\colhead{ID} & \colhead{$V$} & \colhead{$B-V$} & \colhead{$U-B$} & \colhead{$J$} & \colhead{$H$} & \colhead{$K_s$} & \colhead{SpT} & \colhead{$\pi$ (P97)} & \colhead{$\pi$ (V07)} }
\startdata
$\zeta$ Gem	&	3.90	&	0.82	&	--	&	2.38	&	2.03	&	2.14	&	--	&	$	2.79	\pm	0.81	$ &	$	2.37	\pm	0.30	$	\\
HD 49381	&	6.80	&	0.01	&	--	&	6.66	&	6.70	&	6.65	&	A0III	&	$	3.87	\pm	0.96	$ &	$	3.12	\pm	0.47	$	\\
HD 50634	&	6.98	&	-0.02	&	--	&	6.96	&	7.00	&	6.96	&	B9.5II	&	$	2.41	\pm	0.98	$ &	$	3.91	\pm	0.60	$	\\
HD 51354	&	7.12	&	-0.18	&	-0.65	&	7.20	&	7.18	&	7.05	&	B3Vnne	&	$	2.57	\pm	0.81	$ &	$	3.75	\pm	0.47	$	\\
HD 53588	&	7.20	&	-0.11	&	-0.40	&	7.33	&	7.44	&	7.44	&	B7.5IV	&	$	4.67	\pm	0.87	$ &	$	3.96	\pm	0.56	$	\\
HD 51102	&	7.42	&	-0.12	&	--	&	7.56	&	7.66	&	7.70	&	B6V	&	$	0.36	\pm	1.00	$ &	$	1.59	\pm	0.52	$	\\
HD 55919	&	7.43	&	-0.09	&	--	&	7.60	&	7.68	&	7.72	&	B7IV-V	&	$	3.98	\pm	0.79	$ &	$	3.86	\pm	0.60	$	\\
HD 52372	&	7.58	&	0.03	&	--	&	7.42	&	7.45	&	7.43	&	A1III	&	$	1.94	\pm	0.93	$ &	$	2.17	\pm	0.86	$	\\
HD 50767	&	7.70	&	-0.20	&	-0.84	&	8.09	&	8.20	&	8.25	&	B2V	&	$	2.15	\pm	1.01	$ &	$	2.52	\pm	0.74	$	\\
HD 51353	&	7.76	&	-0.07	&	--	&	7.86	&	7.93	&	7.92	&	B7V	&	$	2.00	\pm	0.94	$ &	$	3.22	\pm	0.64	$	\\
HD 52474	&	7.83	&	0.06	&	--	&	7.62	&	7.62	&	7.59	&	A2V	&	$	3.07	\pm	1.02	$ &	$	2.79	\pm	0.66	$	\\
HD 52422	&	7.93	&	-0.01	&	--	&	7.95	&	7.98	&	7.97	&	B9IV	&	$	2.03	\pm	1.10	$ &	$	3.65	\pm	0.86	$	\\
HD 52371	&	8.00	&	-0.07	&	--	&	8.12	&	8.16	&	8.23	&	B7V:	&	$	1.43	\pm	0.97	$ &	$	1.13	\pm	0.83	$	\\
HD 50509	&	8.04	&	-0.08	&	--	&	8.09	&	8.16	&	8.16	&	B8V	&	$	-1.28	\pm	1.06	$ &	$	0.88	\pm	0.72	$	\\
HD 57070	&	8.14	&	-0.04	&	--	&	8.19	&	8.25	&	8.24	&	B9IV	&	$	1.07	\pm	1.05	$ &	$	1.26	\pm	0.66	$	\\
HD 54404	&	8.40	&	-0.05	&	--	&	8.38	&	8.47	&	8.45	&	B9IV	&	$	3.12	\pm	1.05	$ &	$	1.49	\pm	0.86	$	\\
HD 54692	&	8.52	&	0.00	&	--	&	8.49	&	8.54	&	8.53	&	B9.5IV	&	$	3.29	\pm	1.12	$ &	$	2.26	\pm	0.84	$	\\
HD 50107	&	8.84	&	-0.01	&	--	&	8.83	&	8.80	&	8.78	&	--	&		--			&		--				\\
HD 53230	&	8.84	&	0.09	&	--	&	8.59	&	8.54	&	8.51	&	--	&	$	1.92	\pm	1.24	$ &	$	2.56	\pm	1.02	$	\\
HD 53288	&	8.84	&	-0.02	&	-0.08	&	8.81	&	8.85	&	8.84	&	B9.5V	&	$	2.29	\pm	1.05	$ &	$	1.53	\pm	0.83	$	\\
HD 50164	&	8.94	&	-0.07	&	--	&	8.91	&	8.96	&	8.92	&	B9Vnn	&	$	3.82	\pm	1.33	$ &	$	3.34	\pm	1.19	$	\\
HD 49824	&	8.97	&	0.10	&	--	&	8.70	&	8.66	&	8.67	&	--	&	$	0.33	\pm	1.25	$ &	$	-0.74	\pm	0.89	$	\\
HD 263791	&	8.98	&	-0.05	&	--	&	8.99	&	9.12	&	9.06	&	B9V	&	$	0.38	\pm	1.36	$ &	$	0.17	\pm	1.24	$	\\
HD 51187	&	9.02	&	0.02	&	--	&	8.93	&	8.97	&	8.92	&	A0V	&		--			&		--				\\
HD 53473	&	9.16	&	0.02	&	--	&	9.09	&	9.13	&	9.07	&	A0V	&	$	0.92	\pm	1.18	$ &	$	0.30	\pm	0.99	$	\\
HD 51971	&	9.38	&	0.06	&	--	&	9.28	&	9.30	&	9.22	&	A1V	&		--			&		--				\\
TYC 1352-582-1	&	9.60	&	0.12	&	--	&	9.26	&	9.22	&	9.15	&	--	&		--			&		--				\\
BD+18$\degr$1470	&	9.92	&	-0.03	&	--	&	9.92	&	9.97	&	9.98	&	B9.5VpCr-Eu	&		--			&		--				\\
$\zeta$ Gem B	&	11.47	&	0.42: 	&	--	&	10.72	&	10.51	&	10.46	&	F4V	&		--			&		--				\\
07:03:23.1 +20:37:59.5	&	11.71	&	0.38:	&	--	&	11.06	&	10.87	&	10.85	&	--	&		--			&		--				\\
07:04:12.7 +20:30:19.7	&	11.83	&	0.57: 	&	--	&	10.79	&	10.49	&	10.44	&	F6V	&		--			&		--				\\
07:03:38.7 +20:40:59.0	&	12.00	&	0.59:	&	--	&	10.95	&	10.68	&	10.61	&	--	&		--			&		--				\\
07:03:52.6 +20:35:16.3	&	12.34	&	0.54:	&	--	&	11.37	&	11.11	&	11.06	&	F7V	&		--			&		--				\\
07:04:28.6 +20:34:47.3	&	12.37	&	0.53:	&	--	&	11.46	&	11.24	&	11.17	&	--	&		--			&		--				\\
07:04:40.4 +20:35:13.1	&	12.45	&	0.46:	&	--	&	11.66	&	11.51	&	11.47	&	--	&		--			&		--				
\enddata
\tablenotetext{1}{\scriptsize{Stars classified by D. Turner (DAO spectra).}}
\tablenotetext{2}{\scriptsize{The star $\zeta$ Gem B is a spectroscopic binary (\S \ref{s-lm}).  \citet{be07} classified $\zeta$ Gem B as an F3.5V.}}
\tablenotetext{3}{\scriptsize{\citet[][P97]{pe97}, \citet[][V07]{vle07}.}}
\tablenotetext{4}{\scriptsize{\citet{vl07} cite $\pi =2.71\pm0.17$ mas for $\zeta$ Gem.}}
\label{table1}
\end{deluxetable}

At least two classical Cepheids featuring HST parallaxes are cluster members \citep[$\delta$ Cep \& $\zeta$ Gem,][]{ze99,ma12}.  Cluster membership provides a  means to secure independent fundamental parameters ($\log{\tau}$, $E_{B-V}$, $W_{VI_c,0}$, $\log{L_{*}/L_{\sun}}$, $M_V$).  The results shall complement a suite of diverse efforts unified by a common objective to reduce uncertainties associated with $H_0$ in order to constrain cosmological models \citep{fe08,ss10,ge11,ng11,ng12,sm11}.  

\subsection*{{\rm \scriptsize ACKNOWLEDGEMENTS}}
\scriptsize{DM is grateful to the following individuals and consortia whose efforts fostered the research: F. van Leeuwen \& M. Perryman (HIP), F. Benedict (HST), J-C. Mermilliod, \citet{sk10}, 2MASS, M. Mayor, N. Evans, T. Harrison, J. Rosvick, P. Stetson (DAOPHOT), WEBDA \citep{pa08}, DAML \citep{di02}, BSM/AAVSO (T. Krajci, A. Henden, M. Templeton, A. Price), CDS, arXiv, and NASA ADS.  WG is grateful for support from the Chilean Center for Astrophysics FONDAP 15010003 and the BASAL Centro de Astrofisica y Tecnologias Afines (CATA) PFB-06/2007.}

\end{document}